\documentclass[twocolumn]{aastex701}
\usepackage{colortbl} 
\usepackage{amssymb}
\usepackage{enumitem}

\begin{document}

\title{Beyond Cloud-9: The case for discovering more HI-rich starless halos}

\author[orcid=0000-0002-3430-3232,sname='Moreno']{Jorge Moreno}
\affiliation{Department of Physics and Astronomy, Pomona College, Claremont, CA 91711, USA}
\affiliation{Carnegie Observatories, 813 Santa Barbara St., Pasadena, CA 91101, USA}
\email[show]{jorge.moreno@pomona.edu}

\author[0000-0002-2651-7281]{Coral Wheeler}
\affiliation{Department of Physics and Astronomy, California State Polytechnic University, Pomona, Pomona, CA 91768, USA}
\email{cwheeler@cpp.edu}

\author[0000-0002-5908-737X]{Francisco~J. Mercado}
\affiliation{Department of Physics and Astronomy, Pomona College, Claremont, CA 91711, USA}
\affiliation{TAPIR, Mailcode 350-17, California Institute of Technology, Pasadena, CA 91125, USA}
\email{francisco.mercado@pomona.edu}

\author[0000-0003-1848-5571]{M.~Katy Rodriguez Wimberly}
\affiliation{Department of Physics and Astronomy, California State University, San Bernardino, San Bernardino, CA 92407, USA}
\email{maria.wimberly@csusb.edu}

\author[0000-0003-0965-605X]{Pratik~J. Gandhi}
\affiliation{Department of Astronomy, Yale University, New Haven, CT 06520, USA}
\email{pratik.gandhi@yale.edu}

\author[0000-0002-8429-4100]{Jenna Samuel}
\affiliation{Department of Astronomy, The University of Texas at Austin, 2515 Speedway, Stop C1400, Austin, TX 78712-1205, USA}
\affiliation{Cosmic Frontier Center, The University of Texas at Austin, Austin, TX 78712}
\email{jenna.samuel@austin.utexas.edu}

\author[0000-0002-1109-1919]{Robert Feldmann}
\affiliation{Department of Astrophysics, Universität Zürich, Zurich, CH-8057, Switzerland}
\email{feldmann@physik.uzh.ch}

\author[0000-0003-4298-5082]{James~S. Bullock}
\affiliation{Department of Physics and Astronomy, University of Southern California, Los Angeles, CA 90089, USA}
\email{dean@dornsife.usc.edu}

\author[0000-0003-0603-8942]{Andrew Wetzel}
\affiliation{Department of Physics \& Astronomy, University of California, Davis, CA 95616, USA}
\email{awetzel@ucdavis.edu}

\author[0000-0002-9604-343X]{Michael Boylan-Kolchin}
\affiliation{Department of Astronomy, The University of Texas at Austin, 2515 Speedway, Stop C1400, Austin, TX 78712-1205, USA}
\affiliation{Cosmic Frontier Center, The University of Texas at Austin, Austin, TX 78712}
\email{mbk@astro.as.utexas.edu}

\author[0000-0003-3729-1684]{Philip~F. Hopkins}
\affiliation{TAPIR, Mailcode 350-17, California Institute of Technology, Pasadena, CA 91125, USA}
\email{phopkins@caltech.edu}

\begin{abstract}

HI-rich starless halos, should they exist, hold great promise for elucidating dark matter halo structure. Yet realizing this potential demands reliable theoretical predictions for their properties and abundances. Indeed, the recent identification of Cloud-9 as a strong HI-rich starless halo candidate in the nearby universe makes such predictions timely. This Letter examines HI-rich ($M_\mathrm{HI} \geq 10^6\,M_\odot$) starless (isolated/central) halos at $z=0$ across three cosmological simulations: FIREbox, Recal-EAGLE and NIVARIA-LG. All three successfully produce such objects, with $M_\mathrm{HI}$ extending up to $\sim\!1$--$2$ dex above Cloud-9, but with number densities that vary by a factor of $\sim30$. These populations span different regions of the $M_\mathrm{HI}$--$M_\mathrm{gas}$--$M_{200}$ space: NIVARIA-LG produces objects with higher $M_\mathrm{HI}$ and $M_\mathrm{gas}$ values ($\gtrsim 10^7$ and $\gtrsim 10^8\,M_\odot$), while FIREbox predicts they lie within extremely narrow ranges of $M_\mathrm{gas} \sim(1.1$--$1.6)\times10^8\,M_\odot$ and $M_{200} \sim(7.8$--$8.6)\times10^9\,M_\odot$. Recal-EAGLE and NIVARIA-LG exhibit a strong $M_\mathrm{gas}-M_{200}$ correlation, with similar slopes but different normalizations. The simulations predict numerical Cloud-9 analogs -- though similarities in the shapes of their HI column-density profiles may be driven by FAST's modest beam; halving it already reveals differences. Collectively, these inter-simulation discrepancies make a compelling case for discoveries beyond Cloud-9: a statistical sample of well-resolved HI-rich starless halos is needed to discriminate amongst competing predictions.

\end{abstract}


\section{Introduction}\label{sec:intro}
\setcounter{footnote}{0} 

Around $\sim6-7$ months ago, \cite{Anand2025} proclaimed confirmation of Cloud-9 as a bonafide HI-rich starless halo  \citep[for earlier efforts of this kind, see][]{Adams2013}. Originally detected by \cite{Zhou2023} with the Five-hundred-meter Aperture Spherical radio Telescope (FAST), and characterized using follow-up observations with the Very Large Array (VLA) and the Green Bank Telescope (GBT), these authors report an HI-mass of $\sim1.4 \times 10^6 M_\odot$ \citep[see also][]{Karunakaran2024}. Furthermore, their Hubble Space Telescope (HST) observations rule out a stellar component above $10^4 M_{\odot}$
at $\simeq99.5\%$ confidence. Although other such candidates have been propounded, either faint stellar components have been eventually detected \citep[e.g.,][]{Jones2024b,Montes2024,Mitrasinovic2026,Siljeg2026} or they have have been identified as tidal debris associated with neighboring gas-rich massive galaxies \citep[e.g.,][]{Roman2021,Minchin2026}.

Cloud-9 and future kindred discoveries open a powerful baryonic window into the nature of dark matter \citep{Trentham2001}. Unlike their star-endowed counterparts \citep{Bradford2015,Benitez-Llambay2021,Rey2022,Herzog2023}, HI-rich {\it starless} halos are immune to internal star-formation driven feedback, and its effects on the underlying dark matter distribution \citep{Sales2022}. Beyond dark matter constraints, HI-rich {\it starless} halos -- which are likely just under the threshold of galaxy formation -- can also provide unique insights on the physics of galaxy ignition \citep{Moreno2026}.

The goal of this Letter is to make the following inquiries. Do modern cosmological hydrodynamic simulations predict HI-rich starless halo populations, and if so, what are their properties and abundances? As a specific test case, are these simulations able to produce Cloud-9 analogs? If the simulations disagree — with each other or with Cloud-9 — what does that tell us about the challenges ahead for both observations and theory?

\section{Methods}\label{sec:methods}

Although many works study starless halos \citep[e.g.,][]{Sawala2016,Benitez-Llambay2017,Fitts2017,Lee2024,Jeon2025, Doppel2026,HIDES}, they do not report/examine HI-rich objects in detail. For this reason, this Letter only focuses on three cosmological simulations: FIREbox \citep{Feldmann2022}, Recal-EAGLE \citep{TBL2026} and NIVARIA-LG \citep{GarciaBethencourt2026}. 

Table~\ref{table:sims} summarizes the key numerical and physical properties of these three simulations; here we only highlight a few salient differences. FIREbox and Recal-EAGLE employ periodic cosmological boxes with comparable volumes and resolutions (although FIREbox's adaptive softening length for the gas component reaches higher spatial resolution), while NIVARIA-LG uses a smaller zoom-in cosmological simulation constrained around a Local Group environment with slightly lower resolution (their starless halos are centrals, all well outside the virial radii of the massive galaxies). Recal-EAGLE and NIVARIA-LG add self-shielding {\it in post-processing}, using a fitting function from \cite{Rahmati2013}. On the other hand, FIREbox applies a local Sobolev/Jeans-length approximation {\it on the fly}, calibrated against \cite{Rahmati2013} -- see Appendix~C of \cite{FIRE2} for more details. Also, both FIREbox and NIVARIA-LG resolve and capture the clumpy, multiphase structure of the interstellar medium (ISM) explicitly, while Recal-EAGLE relies on a polytropic equation-of-state subgrid prescription \citep{Schaye2008}. Lastly, Recal-EAGLE and NIVARIA-LG employ a density based star-formation prescription (at gas densities $>$ 10 cm$^{-3}$); although the former applies a pressure-based Kennicutt-Schmidt rule while the latter imposes a temperature threshold ($T<1.5\times 10^4$ K). FIREbox requires star-forming gas to be self-gravitating, Jeans-unstable, self-shielding/molecular, with density $n\geq300$ cm$^{-3}$. 

In this paper we only examine simulated starless central halos above an HI-rich threshold of $M_{\rm HI}=10^6 M_\odot$\footnote{Recal-EAGLE adopts a threshold of $10^6 \,h^{-1} M_\odot$ instead. Since FIREbox and NIVARIA-LG have no objects between this value and our $10^6 M_\odot$ threshold, we adopt Recal-EAGLE's threshold for their data for easy comparison with their catalog.}. This particular threshold is an arbitrary choice inspired by Cloud-9 itself, because its HI-rich nature makes it more likely to be observable outside the Local Group with current instruments \citep[see Figure~11 of][]{GarciaBethencourt2026}. This threshold also further ensures that every simulated object we study in this paper is adequately resolved (bottom two rows of Table~\ref{table:sims}). For the sake of simplicity, we focus exclusively on {\it central} halos to minimize environmental effects -- though see Section~\ref{sec:discussion}, where we discuss Cloud-9's isolation status.

\begin{figure*}
    \centering
 \includegraphics[width=1\textwidth]{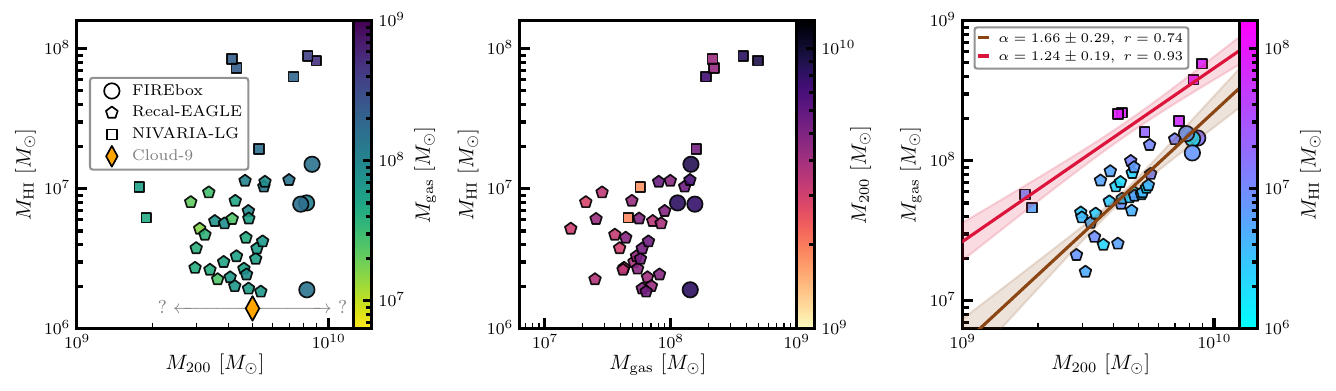}
    \caption{$M_{\rm 200}$-$M_{\rm HI}$ plane (color-coded by $M_{\rm gas}$), $M_{\rm gas}$-$M_{\rm HI}$ plane (color-coded by $M_{\rm 200}$), and $M_{\rm 200}$-$M_{\rm gas}$ plane (color-coded by $M_{\rm HI}$), for the three simulations (FIREbox: circles; Recal-EAGLE: pentagons; NIVARIA-LG: squares). The orange diamond denotes estimates for Cloud-9's halo mass \citep{BLN2023} and HI-gas mass \citep{Anand2025}. The gray arrows with question marks underscore the uncertain assumptions behind Cloud-9's $M_{\rm 200}$ estimate. The brown and red lines and bands are the result of fitting $\log (M_{\rm gas}/M_\odot) = \beta + \alpha \log (M_{\rm 200}/M_\odot)$ for the Recal-EAGLE and NIVARIA-LG samples. We also report Pearson $r$-coefficients for these two populations.
    }
    \label{fig:sample}
\end{figure*}

\section{Results}\label{sec:results}


First we note that the three simulations produce starless HI-rich halos, albeit with distinct abundances. The ``$\#/$Vol" entry in Table~\ref{table:sims} reports the number of objects per unit comoving volume for each simulation. Although FIREbox and Recal-EAGLE have comparable resolutions and volumes, the former predicts over an order of magnitude lower abundance in starless halos ($\sim7\times 10^{-4}$ cMpc$^{-4}$ versus $\sim40 \times 10^{-4}$ cMpc$^{-4}$). Likewise, NIVARIA-LG predicts the highest starless-halo abundance: $\sim200 \times 10^{-4}$ cMpc$^{-3}$.

Figure~\ref{fig:sample} examines the $M_{\rm HI}-M_{\rm gas}-M_{\rm 200}$ space in detail. From left-to-right: projections into the $M_{\rm 200}-M_{\rm HI}$, $M_{\rm gas}-M_{\rm HI}$ and $M_{\rm 200}-M_{\rm gas}$ planes -- color coded by $M_{\rm gas}$, $M_{\rm 200}$ and $M_{\rm HI}$ respectively. The orange diamond in the left panel denotes estimates for Cloud-9's halo mass \citep{BLN2023} and HI-gas mass \citep{Anand2025}. The gray arrows with question marks underscore the uncertain assumptions behind Cloud-9's $M_{\rm 200}$ estimate.

The FIREbox and Recal-EAGLE populations cover similar HI-mass ranges: $M_{\rm HI} \sim(10^6 - 10^7) M_\odot$. On the other hand, more than half of the NIVARIA-LG objects have $M_{\rm HI}\gtrsim 10^7 M_\odot$. Regarding total gas mass, FIREbox has the narrowest range: $M_{\rm gas} \sim (1.1-1.6) \times 10^8 M_\odot$. Meanwhile, the majority of Recal-EAGLE HI-rich starless halos have $M_{\rm gas} \lesssim 1.1\times 10^8 M_\odot$, while most of the NIVARIA-LG objects have $M_{\rm gas} \gtrsim 1.6 \times 10^8 M_\odot$. 

We also note that for Recal-EAGLE and NIVARIA-LG, $M_{\rm gas}$ and $M_{\rm 200}$ scale with one another, with Pearson coefficients $r=0.93$ and $0.74$, respectively. We also fit $\log( M_{\rm gas}/M_\odot) = \beta + \alpha \log (M_{\rm 200}/M_\odot$) for these two simulations and find that the NIVARIA-LG case is slightly steeper and $\sim0.4-0.5$ dex higher than Recal-EAGLE. We do not attempt such an analysis for FIREbox due its small sample size, but note that this case aligns well with the Recal-EAGLE sample on this plane, albeit with higher $M_{\rm 200}$ and $M_{\rm gas}$ values. 

To quantify the differences among these simulations further, Table~\ref{table:w2} reports pairwise Wasserstein-2 distances between the three samples \citep{Villani2009}. With this metric, we find that FIREbox and Recal-EAGLE are the closest pair to each other in $M_{\rm HI}-M_{\rm gas}-M_{\rm 200}$ space, especially along the $M_{\rm 200}-M_{\rm HI}$ projection (Wasserstein-2 distances of $\sim0.6$ and $\sim0.4$ respectively).

\begin{table}
\centering
\caption{Pairwise Wasserstein-2 distances (dex) between the three simulations (FB $=$ FIREbox, EAG $=$ Recal-EAGLE, NIV $=$ NIVARIA-LG),
         computed in each 2D projection and the full 3D space.
         Uncertainties are bootstrap standard deviations (500 resamples).}
\label{table:w2}
\begin{tabular}{lccc}
\hline\hline
Projection & FB--NIV & FB--EAG & NIV--EAG \\
\hline
$\log M_{200}$--$\log M_{\rm HI}$   & $0.87\pm0.16$ & $0.39\pm0.07$ & $0.99\pm0.14$ \\
$\log M_{\rm gas}$--$\log M_{\rm HI}$ & $0.86\pm0.18$ & $0.53\pm0.07$ & $1.13\pm0.17$ \\
$\log M_{200}$--$\log M_{\rm gas}$  & $0.49\pm0.08$ & $0.55\pm0.05$ & $0.57\pm0.10$ \\
\hline
3D & $0.94\pm0.15$ & $0.61\pm0.07$ & $1.15\pm0.16$ \\
\hline
\end{tabular}
\end{table}

\begin{figure}
    \centering
    \includegraphics[width=0.47\textwidth]{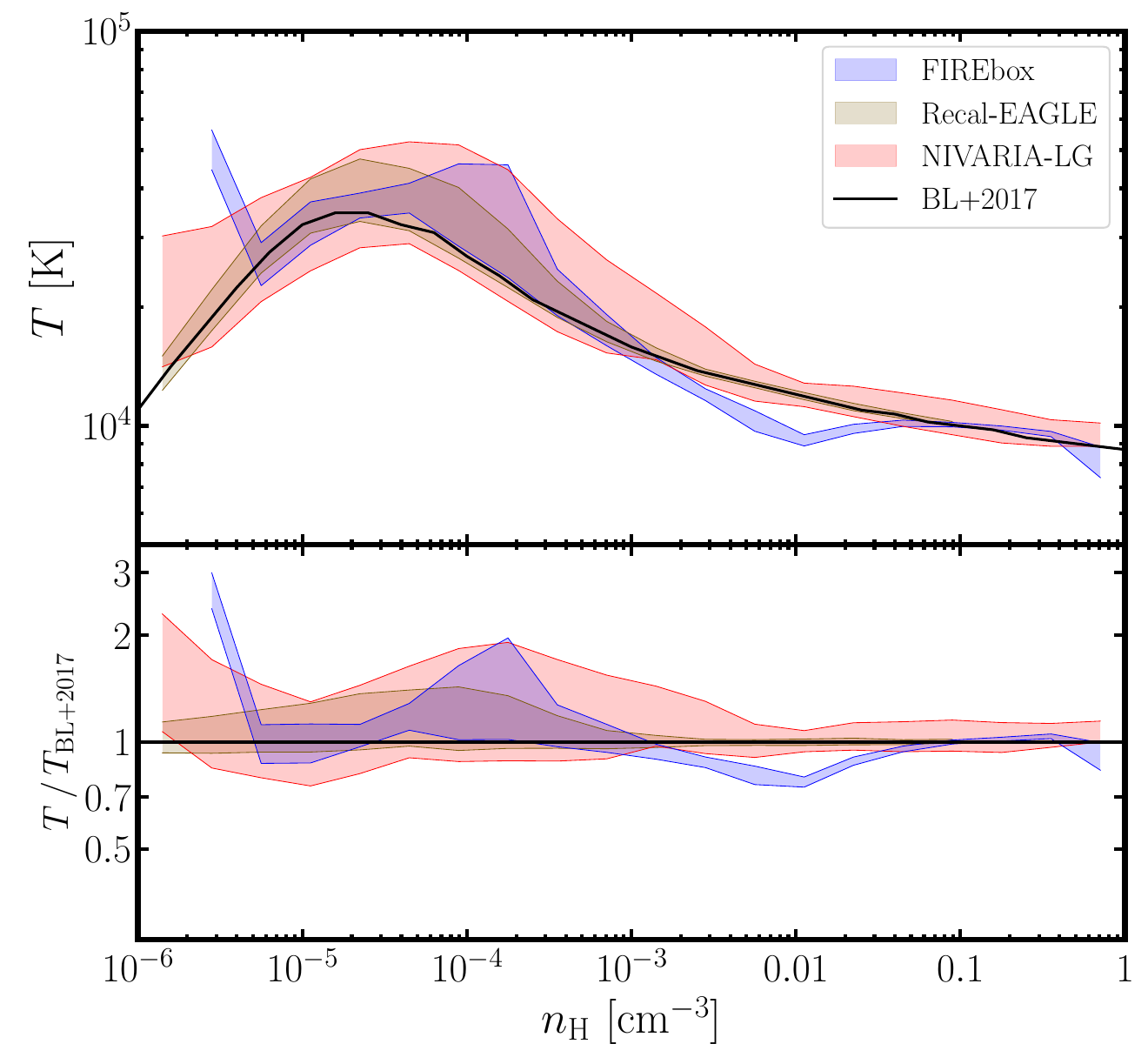}
        \caption{Temperature-density diagram. {\it Top panel}: Bands represent median and inter-quartile behavior of gas particles in HI-rich starless halos selected from FIREbox (blue, 4 objects), Recal-EAGLE (brown, 30 objects) and NIVARIA-LG (red, 8 objects). The solid black line represents the \cite{Benitez-Llambay2017} analytic model. {\it Bottom panel}: Same as above, but normalized to the analytic model.
    }
    \label{fig:phase_diagram}
\end{figure}

Next we explore predictions for the gas-phase structure in simulated HI-rich starless halos. The top panel of Figure~\ref{fig:phase_diagram} shows temperature-density diagrams for the three simulations. The blue, brown and red inter-quartile bands represent FIREbox, Recal-EAGLE and NIVARIA-LG respectively. The black curve denotes the analytic model by \cite{Benitez-Llambay2017}. We also show the three bands modulo this model in the bottom panel. All three simulations exhibit reasonable agreement (within a factor $\lesssim2$) with the analytic model, despite being based on very different physical prescriptions, resolutions and volumes (Table~\ref{table:sims}). 

\begin{figure}
    \centering
     \includegraphics[width=0.47\textwidth]{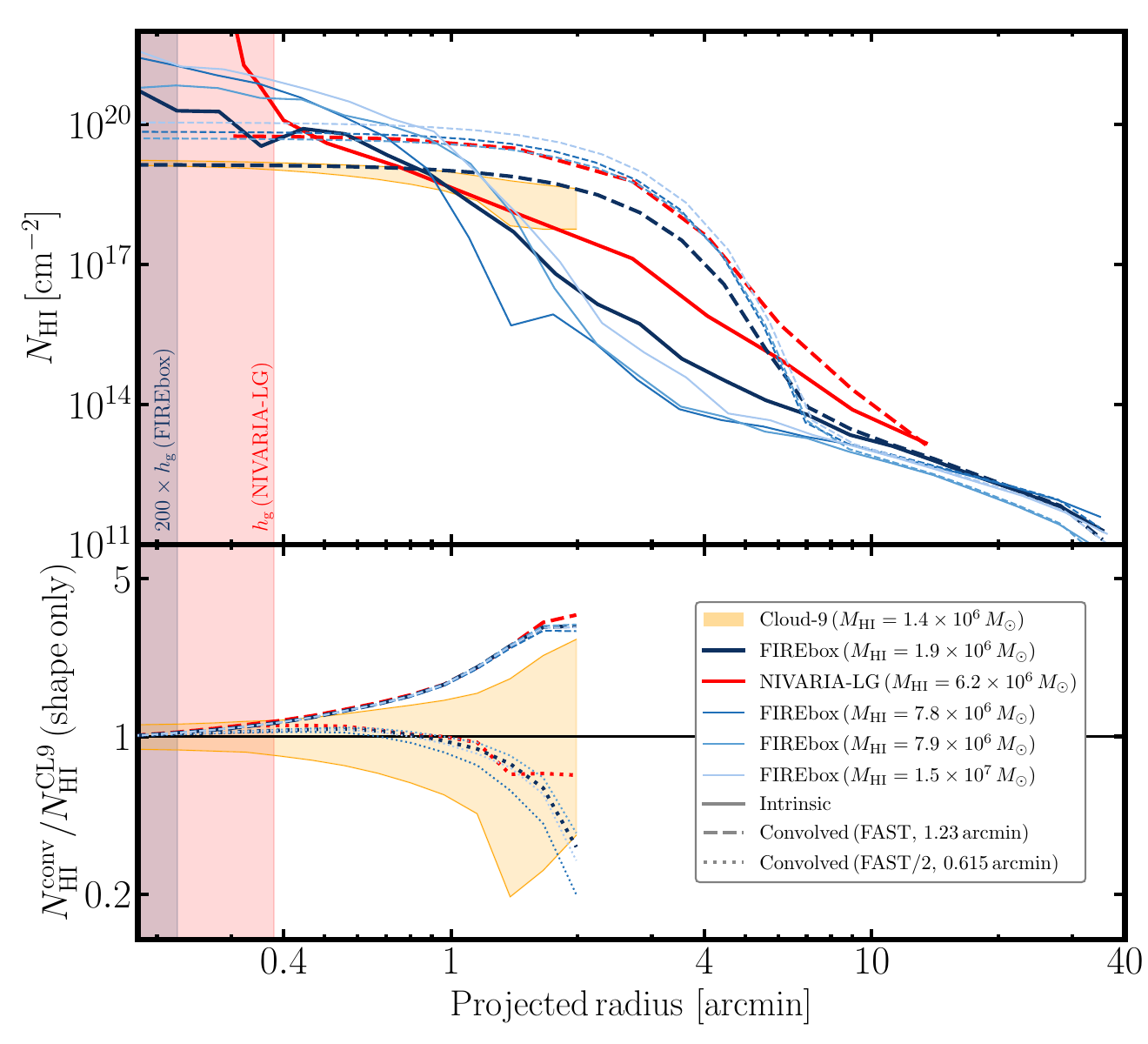}
    \caption{{\it Top panel:} Cloud-9's HI column density profile \citep[orange band,][]{BLN2023}. The solid and dashed curves represent intrinsic and convolved (using a FAST Gaussian beam with $\sigma_{\rm beam}=1.23'$) profiles of simulated halos, placed at at Cloud-9's distance \citep[4.41 Mpc,][]{Anand2021}. The dark-blue and red thick curves represent the objects in FIREbox and NIVARIA-LG with $M_{\rm HI}$ values closest to Cloud-9. The thin lines represent the other three FIREbox objects, with higher $M_{\rm HI}$ values. The simulated profiles extend out to the viral radius. {\it Bottom panel:} The simulated convolved profiles normalized to Cloud-9 observations, and re-normalized to match at the inner most projected-radius bin. The dotted lines show the result of reducing the beam size by a factor of two. The left-hand red (blue) vertical band in both panels denotes NIVARIA-LG's (FIREbox's) gas softening length ($\times 200$). 
    } 
    \label{fig:profiles}
\end{figure}

Lastly we compare the radial structure of Cloud-9's HI component against simulated analogs. The shaded orange region in the top panel of Figure~\ref{fig:profiles} displays Cloud-9's HI column density profile \citep{BLN2023}. We compare against two simulated analogs, one from FIREbox (thick dark-blue) and one from NIVARIA-LG (thick red), chosen to have $M_{\rm HI}$ values as close to Cloud-9 as possible. The thick solid curves show intrinsic predictions, projected onto a random plane (orthogonal projections yield nearly identical results). To compare with observations, the dashed thick lines show the result of convolving with a Gaussian beam following FAST observations \citep[$\sigma_{\rm beam}= 1.23'$,][]{NanFAST}. We display the simulated profiles out to the virial radius.

The blue and red vertical bands represent each simulation's spatial resolution limits. Here we use FIREbox's minimum adaptive length of 1.5 pc, and multiply by a factor of 200, which corresponds to $\sim0.23$ arcmin at this distance. Within the NIVARIA-LG resolution limit ($\lesssim 0.4$) arcmin, the FIREbox object contains 47 gas particles with mean softening length of 80 pc (3.6 arcsec). In other words, both simulated objects are well resolved at projected radii $\gtrsim0.4$ arcmin.

The convolved profile of the FIREbox object matches Cloud-9 reasonably well. The NIVARIA-LG profile is very similar in shape, but with an overall higher normalization. We suspect that this is driven by total HI-content. Concretely, Cloud-9 and the FIREbox object have $M_{\rm HI}=1.4 \times 10^6 M_{\odot}$ and $M_{\rm HI}=1.9 \times 10^6 M_{\odot}$, respectively, while the NIVARIA-LG halo has $M_{\rm HI}=6.2 \times 10^6 M_{\odot}$. To explore the role of total HI mass, we also include the other three FIREbox objects, which have higher $M_{\rm HI}$ values (and thus have better mass resolution) -- which we display as thin curves with blue shades lighter with increasing $M_{\rm HI}$. 

We note that these additional objects also resemble Cloud-9's profile in shape, but not in normalization. To assess similarities in shape, the bottom panel rescales the simulated convolved profiles' normalization to Cloud-9's innermost value, and divide by Cloud-9's profile. Indeed, the five convolved simulated profiles have nearly identical shapes, and are roughly consistent with Cloud-9's shape, especially a small projected radii ($\lesssim0.7$ arcmin).  Note that the shapes of the {\it intrinsic} profiles are not identical to each other, suggesting that similarities among the convolved versions might be driven by FAST's modest beam. To test this, the dotted curves show the result of decreasing the beam size by two -- which now allow us to discern shape differences.

\section{Discussion}\label{sec:discussion}

While the three simulations successfully produce HI-rich starless halos, these populations differ in $M_{\rm HI}-M_{\rm gas}-M_{\rm 200}$ space. Although it might be desirable to conduct direct numerical comparisons, it is not feasible to re-run these cosmological simulations with altered physical assumptions due to their extremely large computational requirements. Rather, we elect to discuss potential reasons behind their differences, and leave carefully-crafted numerical comparisons (utilizing targeted zoom-in simulations as a first step) to future work.

In Figure~\ref{fig:sample}, the NIVARIA-LG population stands out because its starless halos have $M_{\rm gas}$ and $M_{\rm HI}$ values (at fixed $M_{\rm 200}$) higher than the other two simulations. One possible culprit is the fact that this simulation is constrained to a Local Group environment. It would be interesting to compare the NIVARIA-LG results with the original simulations by \cite{Benitez-Llambay2017}, which employ the Recal-EAGLE framework to Local-Group-like environments \citep{Sawala2016b}. Another possibility is NIVARIA-LG's lower resolution (although the objects we present here are fully resolved).  We note that, at the halo mass scale of the four starless halos, FIREbox {\it does} form galaxies in other halos \citep{Feldmann2022} -- with $M_{\star}\sim10^6 M_\odot$, corresponding to $\sim3\times$ the stellar-mass resolution in NIVARIA-LG. Therefore, if resolution effects prevent galaxies with $M_\star \lesssim 10^6~M_\odot$ from forming in these lower resolution simulations, it could explain both the higher gas mass and the relatively higher abundance of HI-rich starless galaxies in that work. These authors plan to release a higher-resolution version of this run (Contreras-Santos, in prep), which would help resolve this question. Given that NIVARIA-LG and Recal-EAGLE's star-formation criteria are similar, we suspect that this mechanism does not play a significant role behind the different outputs from these two simulations.

Likewise, the FIREbox population differs from the other simulations on two fronts: smaller comoving densities and more constrained ranges in $M_{\rm gas}$ and $M_{\rm 200}$. Given its similarity to Recal-EAGLE in terms of volume and resolution (except for the adaptive gas softening length), we suspect that discripancies might be governed by subgrid physics. Concretely, FIREbox includes the following features: explicit treatment of the multi-phase ISM, on-the-fly local Sobolev-Jeans-length self-shielding, and a multi-channel star formation scheme (see Table~\ref{table:sims}). Whether or not a starless halo contains HI is governed by two competing processes: its ability to produce HI and its inability to make stars (plus a possible third: its ability to retain HI gas against external factors). Environmental effects aside, it would be interesting to run dedicated (zoom-in) experiments to explore the role of the aforementioned features in creating HI reservoirs and their consumption into new stars.

In this work we also examine the temperature-density gas structure of our simulated HI-rich starless halos (Figure~\ref{fig:phase_diagram}). We find that the \cite{Benitez-Llambay2017} analytic model exhibits reasonable agreement with the three simulations. Agreement with Recal-EAGLE is expected but not automatic, because that model was originally validated with a Local-Group version of that simulation \citep{Sawala2016b}. Likewise, agreement with NIVARIA-LG is reassuring given that, unlike Recal-EAGLE, this simulation tracks the multi-phase structure of the ISM explicitly. FIREbox also provides reasonable agreement, albeit with only four objects. Although the \cite{Benitez-Llambay2017} analytic model fares reasonably well against three very different simulations, limitations remain. First, that formulation is not capable of predicting the scatter about their $T-n_{\rm H}$ relation. But more importantly, exhibiting agreement with median/inter-quartile behavior does not necessarily guarantee that this model will describe the $T-n_{\rm H}$ plane on a halo-by-halo basis. Expanding that analytic formulation to explain such complexity would make that model more powerful.

We also investigate whether or not our simulated objects are able to reproduce Cloud-9's 21 cm HI column density profile \citep{BLN2023}. We find that one of the FIREbox objects agrees reasonably well with these observations (Figure~\ref{fig:profiles}). The shape of the profile for the NIVARIA-LG object with the lowest $M_{\rm HI}$ also agrees with Cloud-9 -- although the normalization differs (this is expected, given that this object has higher HI-content by a factor of $\sim5$). Nevertheless, the shapes of the intrinsic profiles for these two objects are different, leading us to speculate that the similarities in their convolved counterparts might be driven by FAST's modest Gaussian beam. Indeed, we show that, for the convolved simulated profiles, halving this beam size already uncovers shape differences, underscoring the power of future observations with higher spatial resolution. Surprisingly, characterizations at larger projected radii with FAST's current beam size are unlikely to provide much benefit (i.e., the similarity between convolved profile shapes extends beyond 2 arcmin).

Lastly, although the agreement between the FIREbox object and Cloud-9 is reasonable, the observations systematically exhibit lower HI column density values at larger projected radii. A possible culprit is environment (recall that we only target central halos). A recent study by \cite{TBL2026} points out that estimating $M_{\rm 200}$ from observed HI column density profiles is far from trivial, highlighting the potentially important role played by environment. Indeed, the highly asymmetric nature of the outermost isocontours measured by \cite{BL2024b} with the Very Large Array (VLA) suggest that Cloud-9's structure has been affected by environment, possibly by ram-pressure stripping. Notably, Cloud-9 is located $\sim$104 kpc away \citep[in projection,][]{Zhou2023} from M94 (NGC 4736), a nearby post-starburst spiral with stellar mass $M_\star \sim 5.4\times 10^{10}M_\odot$ \citep{Zhou2023}. In comparison, M94's stellar halo extends out to $\sim 30$ kpc \citep{Gozman2023} and its HI-disk out to $\sim 47$ kpc \citep{Zhou2023}. This galaxy also contains a diffuse HI-filament and seven high velocity clouds (HVCs) within $\sim 50$ kpc of its center, plus two classical satellites at $\sim 38$ kpc and $\sim 69$ kpc \citep[in projection,][]{Smercina2018b}. From abundance-matching considerations \citep{Behroozi2010,Moster2010}, we expect M94's virial radius to be $\sim 250-300$ kpc. In other words, the isolation of Cloud-9 has yet to be firmly established; a satellite or backsplash origin \citep{Diemer2021,Benavides2021,Moreno2022} cannot be ruled out. Indeed, \cite{Zhou2023} suggests that M94 might have experienced recent merging, which are believed to be responsible for tidal debris \citep{Moreno2019}.  This, along with its highly asymmetric shape \citep{BL2024b}, might not fully extricate Cloud-9 from a tidal-debris origin.

\section{Conclusions}\label{sec:conclusions}

The recent detection and characterization of Cloud-9, a candidate HI-rich starless halo in the local universe ($M_{\star} \lesssim 10^{3.5} M_{\odot}$, $M_{\rm HI}\simeq10^6 M_\odot$ and $M_{\rm HI}/M_\star \gtrsim 443$) was no small feat \citep{Anand2025}. As is often the case with exotic/extreme new findings -- e.g., thin satellite planes, galaxies lacking dark matter, the tiniest galaxies, etc. \citep{Ibata2013,vanDokkum2018,Moreno2022,Errani2024u1} -- this particular discovery may pose interesting challenges to simulations. To assess this, in this Letter we compare predictions from three state-of-the-art cosmological simulations: FIREbox \citep{Feldmann2022}, Recal-EAGLE \citep{TBL2026} and NIVARIA-LG \citep{GarciaBethencourt2026}. See Table~\ref{table:sims} for details on the similarities and differences between the simulations. In summary:

\begin{itemize}[noitemsep]
   \setlength{\itemsep}{0pt}
   \setlength{\parsep}{0pt}

    \item The three simulations are capable of producing isolated HI-rich starless halos. $M_{\rm HI}$ predictions extend up to $\sim1-2$ dex above Cloud-9, which is optimal for their potential future detection. However, they generate different comoving number densities at $z=0$: $\sim$($7$, $40$, $200$) $\times 10^{-4}$ cMpc$^{-3}$ (FIREbox, Recal-EAGLE, NIVARIA-LG).
    
    \item The simulations predict different coverage in  $M_{\rm HI}-M_{\rm gas}-M_{\rm 200}$ space (Figure~\ref{fig:sample} and Table~\ref{table:w2}). Salient differences include:
    \begin{itemize}[noitemsep]
        \item[\(\circ\)] NIVIARIA-LG produces objects with higher $M_{\rm HI}$ and $M_{\rm gas}$ values ($\gtrsim10^7 M_\odot$, $\gtrsim10^8 M_\odot$). 
        \item[\(\circ\)] FIREbox produces narrower $M_{\rm gas}$ and $M_{\rm 200}$ ranges: $\sim$($1.1-1.6)\times10^8 M_\odot$ and $\sim(7.8-8.6)\times10^9 M_\odot$.
    \end{itemize}
    
    \item Recal-EAGLE and NIVARIA-LG predict that $M_{\rm gas}$ and $M_{\rm 200}$ are strongly correlated, with similar slopes, $\alpha=(1.66,1.24)$, and Pearson $r-$coefficients $(0.74,0.93)$ -- but different normalizations (NIVARIA-LG $\sim0.4-0.5$ dex higher). FIREbox only has four objects, which closely follow the Recal-EAGLE relation.
    
    \item Regarding the temperature-density phase diagram (Figure~\ref{fig:phase_diagram}), the simulations agree reasonably well (within a factor $\lesssim2$) with the \cite{Benitez-Llambay2017} analytic model.
    
    \item The simulations are also able to create numerical analogs displaying HI column density profiles similar to Cloud-9 \citep{BLN2023}. However, similarities in shape might be driven by FAST's modest beam size \citep[$\sigma_{\rm beam}=1.23'$,][]{NanFAST}. We show that reducing the beam size by a factor of two already reveals differences across profile shapes (Figure~\ref{fig:profiles}).

\end{itemize}

Despite the aforementioned intricacies, HI-rich starless halos are simpler than galaxies because they do not suffer from the effects of internal star formation and its associated feedback \citep{BullockMBK2017,Sales2022}, allowing us in principle to marginalize over baryonic processes to constrain the nature of dark matter \citep{Tulin2018,Fitts2019}. Moreover, this population can further enlighten our understanding of galaxy ignition \citep{Moreno2026}. However, in order to achieve this, it is imperative to detect and characterize more such systems. Indeed, the three simulations we study here make distinct predictions for the abundance and properties of HI-rich starless halos — differences that current observations of a single object are unable to adjudicate.

For these reasons, we urge the community to seek discoveries beyond Cloud-9 and allocate resources toward both new HI-rich starless halos and dedicated follow-up of existing candidates in the Local Universe \citep[e.g.,][]{Monaci2026} -- and to undertake this with higher spatial resolution than currently available. Along this vein, our findings suggest that reducing FAST's beam size by only a factor of two already reveals differences in the HI column density profile shapes that are otherwise washed out by the current beam. Indeed, the discovery of a statistical sample of well-resolved HI-rich starless halos would serve a dual purpose: opening a new window onto the physics of galaxy formation and dark matter, while simultaneously providing the observational leverage needed to discriminate amongst the simulation predictions presented here. Beyond this, it is key that such future investigations also scrutinize the purported absence (or presence) of a faint stellar component — as well as their isolation status — via detailed spectroscopic observations and kinematic measurements. \newpage

\begin{table*}[t]
\centering
\caption{Key properties of the three simulations used in this work. $m$, $h$, $n$ and $T$ refer to particle mass (before mass loss for stellar particles) and therefore the mass resolution for different species, softening length (proxy for spatial resolution), number density and temperature respectively. LG $=$ Local Group. MFM $=$ Meshless finite mass. SPH $=$ Smoothed particle hydrodynamics. UVB $=$ Ultraviolet background. $z_\mathrm{re}=$ reionization redshift (different phases). ISM $=$ interstellar medium. EOS $=$ Equation of state. SF $=$ Star formation. $\alpha_{\rm vir}=$ virial parameter. SFE $=$ Star formation efficiency. KS $=$ Kennicut-Schmidt \citep{Kennicutt}. $\#$ and $\#/{\rm Vol}$ denote the number and comoving number density of HI-rich starless halos at $z=0$ (with $M_{\rm HI} > 10^6 M_{\odot}$), while the last two rows refer to the gas and dark-matter particle numbers of the {\it least} resolved objects.}
\label{tab:fullwidth}
\begin{tabular}{cccc}
\hline \hline
 & \textcolor{blue}{\textbf{FIREbox}} & \textcolor{brown}{\textbf{Recal-EAGLE}} & \textcolor{red}{\textbf{NIVARIA-LG}}\\ 
 & \cite{Feldmann2022} & \cite{TBL2026} & \cite{GarciaBethencourt2026} \\
\hline
Type      & Cosmological volume   & Cosmological 
volume        & Cosmological zoom-in  \\       &   (uniform / periodic)  &  
 (uniform / periodic)        &  (constrained LG) \\
Hydro                 & \textsc{gizmo} (meshless MFM)        & \textsc{p-gadget3} (SPH)             & \textsc{gasoline2} (SPH) \\
Physics                 & FIRE-2        & EAGLE            & NIHAO \\
                 & \citep{FIRE2} & \citep{Schaye2015}            & \citep{Wang2015} \\
Volume             & $L = 22.1$~cMpc (periodic box)                      & $L = 20$~cMpc (periodic box)                  & $r = 7.4$~cMpc (zoom-in sphere)\\
\arrayrulecolor{lightgray}\hline\arrayrulecolor{black} 
$m_{\rm dm}$    & ${\sim}3.5\times10^{5}$~$M_\odot$   & $2.8\times10^{5}$~$M_\odot$          & $1.6\times10^{6}$~$M_\odot$ \\
$m_{\rm gas}$   & ${\sim}6.3\times10^{4}$~$M_\odot$   & $5.26\times10^{4}$~$M_\odot$         & $3\times10^{5}$~$M_\odot$ \\
$m_{\star}$ & ${\sim}6.3\times10^{4}$~$M_\odot$ & ${\sim}5.26\times10^{4}$~$M_\odot$ & ${\sim}3\times10^{5}$~$M_\odot$ \\
\arrayrulecolor{lightgray}\hline\arrayrulecolor{black} 
$h_\mathrm{dm}$   & 80~pc (Plummer equivalent)  & ${\sim}195$~pc & 860~pc \\
$h_\mathrm{gas}$ & minimum 1.5~pc (adaptive; & ${\leq}195$~pc  & 488~pc \\
 & inter-particle spacing & (${\sim}1\%$ of mean &  \\
 & at SF threshold ${\sim}20$~pc) & inter-particle separation) &  \\
$h_\star$    & 12~pc (Plummer equivalent) & ${\sim}195$~pc & ${\sim}488$~pc \\
\arrayrulecolor{lightgray}\hline\arrayrulecolor{black} 
UVB & $z_\mathrm{re} \approx 10$ (onset) &  $z_\mathrm{re} = 11.5$ (instantaneous) & $z_\mathrm{re} \approx 6$ (completion) \\
 &  \citep{Faucher2009}   & \citep{HaardtMadau2012} & \citep{HaardtMadau2012}; \\
\arrayrulecolor{lightgray}\hline\arrayrulecolor{black} 
Self-       & Local Sobolev/Jeans-length  & \citet{Rahmati2013} & \citet{Rahmati2013}  \\
shielding                     & approximation calibrated against  & fitting function  & fitting function  \\
                     & \citet{Rahmati2013} and  & applied in post-processing & applied in post-processing \\
                     &  \citet{FaucherGiguere2010}  &  based on radiative transfer &  based on radiative transfer \\
                     &  computed on-the-fly; & of \citet{ Pawlik2008} & of \citet{ Pawlik2008}  \\
                     &  molecular fraction for SF follows &  and \citet{Pawlik2011}&  and \citet{Pawlik2011} \\
                     &  \citet{Krumholz2011}  &  &  \\
\arrayrulecolor{lightgray}\hline\arrayrulecolor{black} 
ISM &  Explicit multiphase: & Subgrid EOS (polytropic)  & Explicit multiphase: \\
     &  hot, warm, and cold phases  & \citep{Schaye2008}  & Cold and warm/hot \\
     &  (down to ~10 K) emerge & applied to SF gas; & phases coexist and\\
     &  from feedback-regulated  & cold ISM not resolved & are tracked directly \\
     & FIRE-2 physics & above SF threshold & \\
\arrayrulecolor{lightgray}\hline\arrayrulecolor{black} 
SF-          & Gas must simultaneously satisfy:  & $n_\mathrm{H} > n_\mathrm{th} = 10$~cm$^{-3}$ & $n_\mathrm{H} > n_\mathrm{th} = 10$~cm$^{-3}$ \\
criteria          & (1) self-gravitating ($\alpha_\mathrm{vir} < 1$)  & pressure-based KS & and $T < 1.5\times10^{4}$~K\\
          & (2) Jeans-unstable  &  & \\
          & (3) self-shielding/molecular  &  &\\
          & (4) $n \geq 300$~cm$^{-3}$;  &  &\\
          & SFE $= 100\%$ per local free-fall &  &\\
          & applied to molecular gas only &  &\\
          & \citep{FIRE} &  &\\
\arrayrulecolor{lightgray}\hline\arrayrulecolor{black} 
$\#$ & 4 & 30 & 8 \\
 $\#/{\rm Vol}$ & $\approx 7.4 \times 10^{-4}  \, {\rm cMpc}^{-3}$ & $\approx4.1 \times 10^{-3}  \, {\rm cMpc}^{-3}$ &$\approx2.0 \times 10^{-2}  \, {\rm cMpc}^{-3}$ \\
 min $N_{\rm gas}$ & 2,295  & 299 & 154\\
 min $N_{\rm dm}$  & 21,902 & 8,708 & 1,109\\
\hline
\end{tabular}
\label{table:sims}
\end{table*}


\begin{acknowledgments}
JM is funded by a Pomona College Large Research Grant and thanks the Carnegie Observatories and Pomona's Humanities Studio for their hospitality. FJM is funded by the National Science Foundation (NSF) Math and Physical Sciences (MPS) Award AST-2316748. MKRW acknowledges support from NSF MPS Ascending Faculty Catalyst Award AST-2444751. JBS, JS and MBK acknowledge support for program JWST-AR-06278 by NASA through a grant from the Space Telescope Science Institute, which is operated by the Association of Universities for Research in Astronomy, Inc., under NASA contract NAS 5-03127. RF acknowledges financial support from the Swiss National Science Foundation (grant no 200021-188552). JS and MBK acknowledge support for program JWST-AR-06278 by NASA through a grant from the Space Telescope Science Institute, which is operated by the Association of Universities for Research in Astronomy, Inc., under NASA contract NAS 5-03127. MBK also acknowledges support from NSF grants AST-1910346, AST-2108962, and AST-2408247; NASA grant 80NSSC22K0827; HST-GO-16686, HST-AR-17028, HST-AR-17043, and JWST-GO-03788, from STScI; and from the Samuel T. and Fern Yanagisawa Regents Professorship in Astronomy at UT Austin. AW received support from NSF, via CAREER award AST-2045928 and grant AST-2107772. Support for PFH was provided by a Simons Investigator Award. JM thanks S.~Kim and K.~Nyland for illuminating discussions on an earlier draft -- plus G.~García-Bethancourt, A.~Di Cintio, F. Turini and A.~Benitez-Llambay for support with their data. We thank the anonymous reviewer for their insightful comments. The authors also acknowledge the labor by the cleaning and the clerical staff, the food service workers, the technical support personnel, and many others that make the astronomy research enterprise possible. We conducted this work on Tongva-Gabrielino land.\\
\end{acknowledgments}

\begin{contribution}
JM wrote the entire manuscript and led every aspect of the project. All authors contributed substantially to the final polishing of this manuscript. CW, FJM, and MKRW made significant contributions to the original design of the paper. PJG, JS, RF, JBS, AW, MBK and PFH provided additional support on the big-picture context of the project and the creation of our physics model, while RF created FIREbox. All authors contributed substantially to the final polishing of this manuscript.

\end{contribution}

\textit{\large Data availability}: \url{https://jorgemorenosoto.github.io/HI-rich-starless-halos/}

\software{yt \citep{Turk2011}, Amiga Halo Finder \citep{Knollmann2009}, PlotDigitizer (\url{https://plotdigitizer.com/app)}.}

\section*{Large Language Model Usage Disclosure}\label{sec:llm}

JM utilized and closely supervised Anthropic's Claude to help polish the Figures and some of the writing in this Letter, and takes full responsibility of its outputs.

\newpage
\bibliography{bibliography}{}
\bibliographystyle{aasjournalv7}

\end{document}